\documentclass[sigconf]{acmart}
\usepackage{fancyhdr}
\usepackage{bbding}
\usepackage{amsmath}
\usepackage{graphicx}
\usepackage{subfigure}
\usepackage{multirow}
\usepackage{underscore}
\usepackage{float}
\usepackage{booktabs}

\AtBeginDocument{%
  \providecommand\BibTeX{{%
    \normalfont B\kern-0.5em{\scshape i\kern-0.25em b}\kern-0.8em\TeX}}}

\setcopyright{acmlicensed}
\copyrightyear{2024}
\acmYear{2024}
\setcopyright{acmlicensed}
\acmConference[MRAC '24] {Proceedings of the 2nd International Workshop on Multimodal and Responsible Affective Computing}{November 1, 2024}{Melbourne, VIC, Australia.}
\acmBooktitle{Proceedings of the 2nd International Workshop on Multimodal and Responsible Affective Computing (MRAC '24), November 1, 2024, Melbourne, VIC, Australia}
\acmDOI{10.1145/3689092.3689415}
\acmISBN{979-8-4007-1203-6/24/10}

\acmSubmissionID{mrac0648}


\graphicspath{{./images/}}  
\begin{document}
\pagestyle{fancy}

\begin{sloppypar}
\title{Early Joint Learning of Emotion Information Makes
MultiModal Model Understand You Better}

\author{Mengying Ge}
\authornotemark[1]
\email{gemengying@kanzhun.com}

\affiliation{%
  \institution{BOSS ZhiPin}
  \city{Beijing}
  \country{China}
}

\author{Mingyang Li}
\authornote{Mengying Ge and Mingyang Li contributed equally to this work.}
\email{limingyang01@kanzhun.com}
\affiliation{%
  \institution{BOSS ZhiPin}
  \city{Beijing}
  \country{China}
}

\author{Dongkai Tang}
\email{tangdongkai@kanzhun.com}
\affiliation{%
  \institution{BOSS ZhiPin}
  \city{Beijing}
  \country{China}
}
\author{Pengbo Li}
\email{lipengbo@kanzhun.com}
\affiliation{%
  \institution{BOSS ZhiPin}
  \city{Beijing}
  \country{China}
}
\author{Kuo Liu}
\email{liukuo@kanzhun.com}
\affiliation{%
  \institution{BOSS ZhiPin}
  \city{Beijing}
  \country{China}
}
\author{Shuhao Deng}
\email{dengshuhao@kanzhun.com}
\affiliation{%
  \institution{BOSS ZhiPin}
  \city{Beijing}
  \country{China}
}
\author{Songbai Pu}
\email{pusongbai@kanzhun.com}
\affiliation{%
  \institution{BOSS ZhiPin}
  \city{Beijing}
  \country{China}
}
\author{Long Liu}
\affiliation{%
  \institution{BOSS ZhiPin}
  \city{Beijing}
  \country{China}
}
\author{Yang Song}
\affiliation{%
  \institution{BOSS ZhiPin}
  \city{Beijing}
  \country{China}
}
\author{Tao Zhang}
\affiliation{%
  \institution{BOSS ZhiPin}
  \city{Beijing}
  \country{China}
}

\renewcommand{\shortauthors}{Mengying Ge et al.}


\begin{abstract}

In this paper, we present our solutions for emotion recognition in the sub-challenges of Multimodal Emotion Recognition Challenge (MER2024). For the tasks MER-SEMI and MER-NOISE, participants are required to recognize discrete emotions. Particularly, in MER-NOISE, the test videos are corrupted with noise, necessitating the consideration of modality robustness. We developed our Emotion ViT based on large-scale data pre-training and fine-tune, a vision feature extractor adapted to emotion recognition tasks. In addressing the modal competition between audio and text, we implemented an early fusion methodology underpinned by a large language model. This design facilitates full interaction between audio and text, thereby harmonizing their contributions. The joint Audio-Text representation can be late-fused with other features extracted from our specific unimodal encoders. To solve the problems of data insufficiency and class imbalance, we employed multiple rounds of multi-model voting for data mining. To ensure the high quality of audio features, we introduce a speech source separation method to denoise the audios.Our model secured \textbf{2nd} in both MER2024-SEMI and MER2024-NOISE categories, affirming that the robustness and validity of our strategies in advancing the field of multimodal emotion recognition.
\end{abstract}


\maketitle
\section{Introduction}
In the rapidly advancing field of human-machine interaction, emotion recognition stands as a pivotal bridge facilitating emotional communication between humans and machines. Previous works cues from facial expressions, vocal intonations, and textual content to deeply analyze the emotional states of individuals\cite{picard2000affective}. Among these approaches, multimodal technology with its capability to integrate different modal information sources for comprehensive analysis, has garnered significant attention\cite{hu2022unimse}\cite{hazarika2020misa}. However, the emotional complexity, scarcity of high-quality annotated data and inevitable environmental noise in practical applications pose substantial challenges to the accuracy and robustness of multimodal emotion recognition. Efficient and precise feature extractors have emerged as a crucial means to overcome these bottlenecks. Traditional handcrafted features and machine learning classifiers often falter when confronted with complex nonlinear relationships, while the rise of deep learning and large-scale pre-trained models has infused new vitality into the domain of emotion recognition, exemplified by models such as VideoMAE-large\cite{tong2022videomae}, CLIP-large \cite{radford2021learning} for vision, wav2vec2.0 \cite{baevski2020wav2vec}, HUBERT\cite{hsu2021hubert} for audio, and RoBERTa\cite{liu2019roberta}, Baichuan-13B\cite{yang2023baichuan} for text, which have all demonstrated remarkable performance. Although training of feature extractors and classifiers simultaneously can enhance results further, the large scale parameters in multimodal contexts typically entail prohibitive computational costs. Hence, a more efficient approach involves optimizing feature extractors during pre-training and subsequently freezing their parameters while only training fusion layers and classifier layers\cite{siriwardhana2020jointly}.

Within the realm of multimodal information processing, the information from different modalities complements one another yet may also contain redundancy or conflicts. Effectively fusing these multiple sources of information remains a core challenge in the field\cite{baltruvsaitis2018multimodal}\cite{wang2022systematic}. Notably, late fusion strategies incorporating attention mechanisms have shown exceptional performance in semantic-level feature integration\cite{lian2024merbench}. Nonetheless, it is often observed in practice that even with improvements in individual modality models, the performance of fused models may fall short of expectations, potentially due to modality conflicts, overfitting of unimodal models, and an overreliance on a single modality at the expense of exploiting complementary advantages across modalities\cite{huang2022modality}. Thus, exploring effective joint training strategies to optimize multimodal feature representations becomes imperative.

The MER-SEMI and MER-NOISE tracks specifically address the challenges of semi-supervised learning in multimodal emotion recognition and enhancing robustness under noisy conditions. The organizers have generously provided meticulously annotated high-quality datasets and abundant unlabeled data resources, laying a solid foundation for participants and researchers to delve into these critical areas. Additionally, comprehensive research materials have been released, offering invaluable insights into understanding and addressing these challenges\cite{lian2024mer}. Initially, extensive baseline experiments were conducted to systematically assess the performance of unimodal models in emotion recognition tasks, conclusively demonstrating the substantial performance boost from more powerful feature extractors. Subsequently, in pursuit of further exploiting the potential of multimodal information, feature-level late fusion strategies were attempted to consolidate features from different modalities and enhance overall recognition capabilities. However, reliance solely on late fusion strategies often fails to adequately address potential conflicts and redundancies among modalities, thereby limiting the optimization of fusion effectiveness. In response to this challenge, we propose a novel joint training framework, particularly focusing on early fusion exploration of speech and text modalities. 

This framework incorporates attention mechanisms and employs a hybrid fusion approach to handle multimodal features with greater finesse, effectively alleviating modality conflicts, and fully leveraging the complementary strengths between modalities, ultimately realizing efficient and precise integration and recognition of multimodal emotion features. Our key contributions are summarized as follows: 

\textbf{Emotion ViT Model}: We have trained a Vision Transformer (ViT) model proficient in characterizing human emotional features. Leveraging self-supervised learning, this model efficiently harnesses vast amounts of unlabeled data during pre-training and, through subsequent fine-tuning on a broad range of emotional data, significantly enhances its capacity for emotion feature expression.

\textbf{Audio-Text Joint Training Architecture}: To tackle potential expression conflicts in multimodal emotion recognition, we introduce an innovative joint training structure for speech and text. By adopting early fusion, this structure effectively integrates information from both speech and text modalities, circumventing information loss associated with late fusion and surpassing simplistic dual-modal late fusion methods in performance.

\textbf{Noise Robustness Enhance}: Addressing the interference of noise in complex environments on emotion recognition performance, audio denoise strategies and optimizations to ASR system's noise resilience have been implemented. These measures collectively elevate the model's stability and recognition precision in noisy environments, thereby reinforcing its noise robustness.

\textbf{Efficient Utilization of Unlabeled Data}: To fully exploit the potential of unlabeled data, we incorporate a cyclic boosting data mining method. Through iterative utilization of unlabeled data for model training, this approach significantly enhances model accuracy and generalization capabilities, achieving efficient utilization of unlabeled data.

\section{Proposed Method}
\begin{figure}[h]
  \centering
  \includegraphics[width=\linewidth]{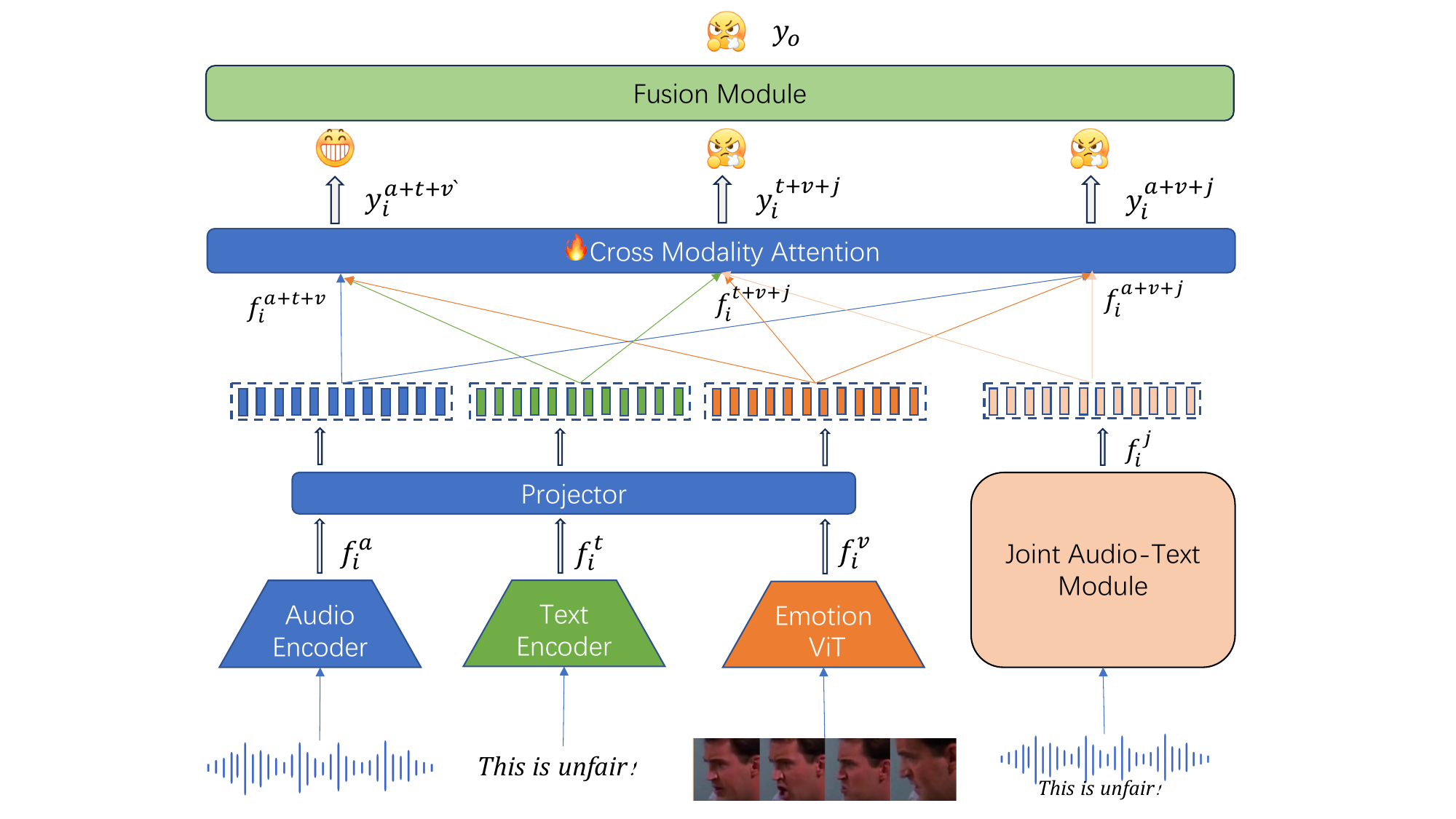}
  \caption{Multi-Modal Emotion Recognition Framework}
\end{figure}
\subsection{Framework Overview}
As shown in Figure 1, we employ audio, text, and vision encoders to extract features from input audio, text, and facial sequence data, respectively. These encoders have been pre-trained or fine-tuned on emotion datasets. Further, we introduce a novel joint Audio-Text modality feature extraction module, specifically designed to harness the synergistic information between audio and text, thereby enhancing the expressivity of cross modalities.

Subsequently, we leverage the distinct characteristics of these extracted features by combining them based on their differential properties. A cross-modal attention mechanism is then employed to fuse these combined features, allowing the system to dynamically weigh and focus on the most salient aspects for emotion recognition across each modality. Each fusion branch yields its independent prediction.
Finally, a tailored ensemble strategy is invoked to ascertain the ultimate predicted result.

As shown in formulas (1)-(7) , where \(f^{t}_{i}\), \(f^{a}_{i}\), and \(f^{v}_{i}\) respectively represents the features extracted from text, audio, and visual modalities.
\begin{gather}
f^{a+t+v}_{i} = f^{t}_{i} \oplus f^{a}_{i} \oplus f^{v}_{i} \\
f^{a+v+j}_{i} = f^{a}_{i} \oplus f^{v}_{i} \oplus f^{j}_{i} \\
f^{t+v+j}_{i} = f^{t}_{i} \oplus f^{v}_{i} \oplus f^{j}_{i} \\
Z^{*+}_{i} = W^{*+} \cdot f^{*+}_{i} + b_{z} \\
p^{*+}_{i} = \text{Softmax}(W^{\text{smax}} Z^{*+}_{i} + b^{\text{smax}}) \\
y^{*+}_{i} = {\arg\max}_(p^{*+}_{i}) \\
y_{o} = \textbf{Ensemble}({y}^{a+t+v}_{i}, {y}^{a+v+j}_{i}, {y}^{t+v+j},...)
\end{gather}
\(f^{j}_{i}\) indicates the early joint Audio-Text modality feature, and \(f^{*+}_{i}\) signifies the fused features based on multimodal attention of different feature groups. Each \(y_{i}\) signifies the prediction outcome from individual fusion branch. \(\hat{y}_{o}\) signifies the final sentiment prediction after our specific ensemble strategy. 

\subsection{Unimodal Emotional Encoder}
\subsubsection{\textbf{Emotion ViT}}
\ 
\newline
\noindent 
For MER2024, a large amount of labeled emotion data is not available. Considering this limitation, we adopt a self supervised training approach and propose EmotionViT, which mainly includes two parts:
\begin{itemize}
\item ViT-Base\cite{dosovitskiy2020image}: We employ the MAE\cite{he2022masked} self-supervision methodology for pre-training,utilizing ViT-Base as the backbone architecture. Our training dataset comprises a substantial 8 million images, including facial recognition, facial attributions, facial occlusion, and facial expression datasets. In particular, we have trained the ViT from scratch without leveraging any publicly available pre-trained models, such as models pre-trained on ImageNet1K\cite{deng2009imagenet}. We believe that although ImageNet1K covers most scenarios, facial expressions are the most important basis for emotion recognition. Therefore, our entire dataset for pre-training contains a large amount of facial data. After pre-trained, the ViT-Base model was fine-tuned on the labeled data of MER2024 and public emotion recognition datasets. Add a classification head to the vit base structure for training facial expression classification, and finally use the fine-tuned model to extract facial features as input for subsequent model fusion. During the process of creating data, we found that, The dataset for fine-tuning comes from the training set of MER2024, which consists of segments of video with video level labels instead of image level labels. For example, the annotated video is labeled as happy, but after the video is divided into frames, although the main expression is happy, not every frame is happy. Therefore, when this data is added to the training, it becomes a negative sample. Based on this, we innovatively distinguish the training data. For the source of a single image, we still use the original input method, as shown in Figure 2 (a), while for the training data of a video, we change the input of vit base from a single image to multiple frames from the same video concatenated together. The resulting grids picture is shown in Figure 2 (b), and this modification can greatly improve the final accuracy of the model and obtain better feature representation capabilities.
\begin{figure}[H]
\centering
\subfigure[]{
\includegraphics[height=5cm]{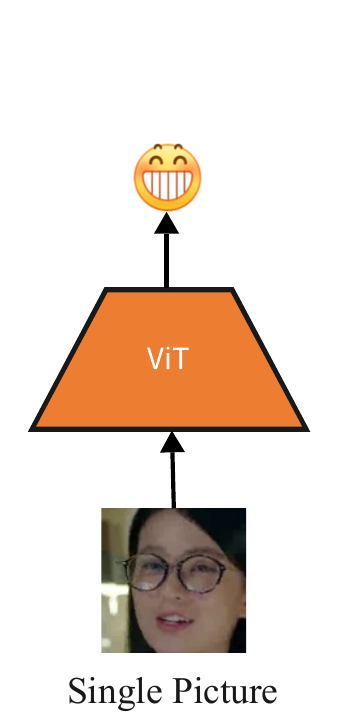}
}
\quad
\subfigure[]{
\includegraphics[height=5cm]{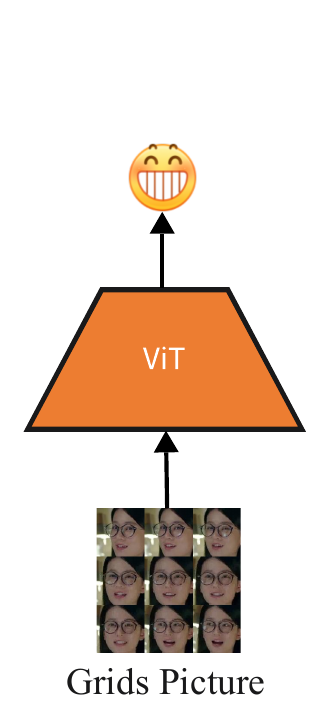}
}
\caption{ViT-Base fine-tune of different inputs}
\end{figure}
\item InternVIT-6B: In addition to self-supervised pre-training methods such as MAE, the CLIP leverages image-text pairs for contrastive learning, is widely adopted in multi-modal models as a visual encoder. For our visual feature fusion, we introduced vit-base pre-trained model and CLIP pre-trained model for feature extraction. The InternVL-chat-V1.5\cite{chen2024internvl}, outstanding on the opencompass multi-modal benchmark, employs InternVIT-6B, a visual encoder pre-trained on a large-scale, high-quality image-text dataset with a substantial 6 billion parameters, exhibiting exceptional feature representation capabilities. We enhanced InternVIT-6B with a classification head and fine-tuned it on labeled data of MER2024 and public emotion datasets. Given the large scale parameters of InternVIT-6B, we only trained the projector layers. 
\end{itemize}
In summary, EmotionViT mainly includes ViT-Base and InternViT-6B, which are pre trained and fintuned on a large amount of human centered data and integrated for use on the MER-SEMI and MER-NOISE tracks.Then the facial frames from a video is transformed into a sequence of high-dimensional video embeddings through Emotion ViT, denoted as \( \mathbf{F}_v \in \mathbb{R}^{N_v \times 768} \).

\subsubsection{fine-tuned Chinese\_HuBERT\_large}
\ 
\newline
\noindent To extract features from raw audio inputs, we employ a fine-tuned variant of the large-scale Chinese version of {HuBERT\_large}. This model has been meticulously adapted to the MSA and Multimodal Emotional Recognition 2023 Audio Datasets, enhancing its capability to comprehend nuanced linguistic and emotional content in the Chinese language. The feature extraction process harnesses the concatenated hidden representations averaged over the last four layers of the {Chinese\_HuBERT\_large} architecture. Consequently, each audio input is transformed into a sequence of high-dimensional audio embeddings, denoted as \( \mathbf{F}_a \in \mathbb{R}^{N_a \times 1024} \).
\subsubsection{fine-tuned BaiChuan13b\_Chat}
\ 
\newline
\noindent We extract features for a text input by leveraging a large language model Baichuan13b\_Chat, fine-tuned on 200w Chinese texts related to emotion which contains WeiBo, ECB and MER2023\cite{lian2023mer} Dataset, and use the averaged hidden representations across the last four layers of the LLM, resulting in a sequence of text representations  \( \mathbf{F}_t \in \mathbb{R}^{N_t \times 5120} \) for each text input.

\subsection{Joint Audio-Text Module}
\begin{figure}[H]
  \centering
  \includegraphics[width=\linewidth]{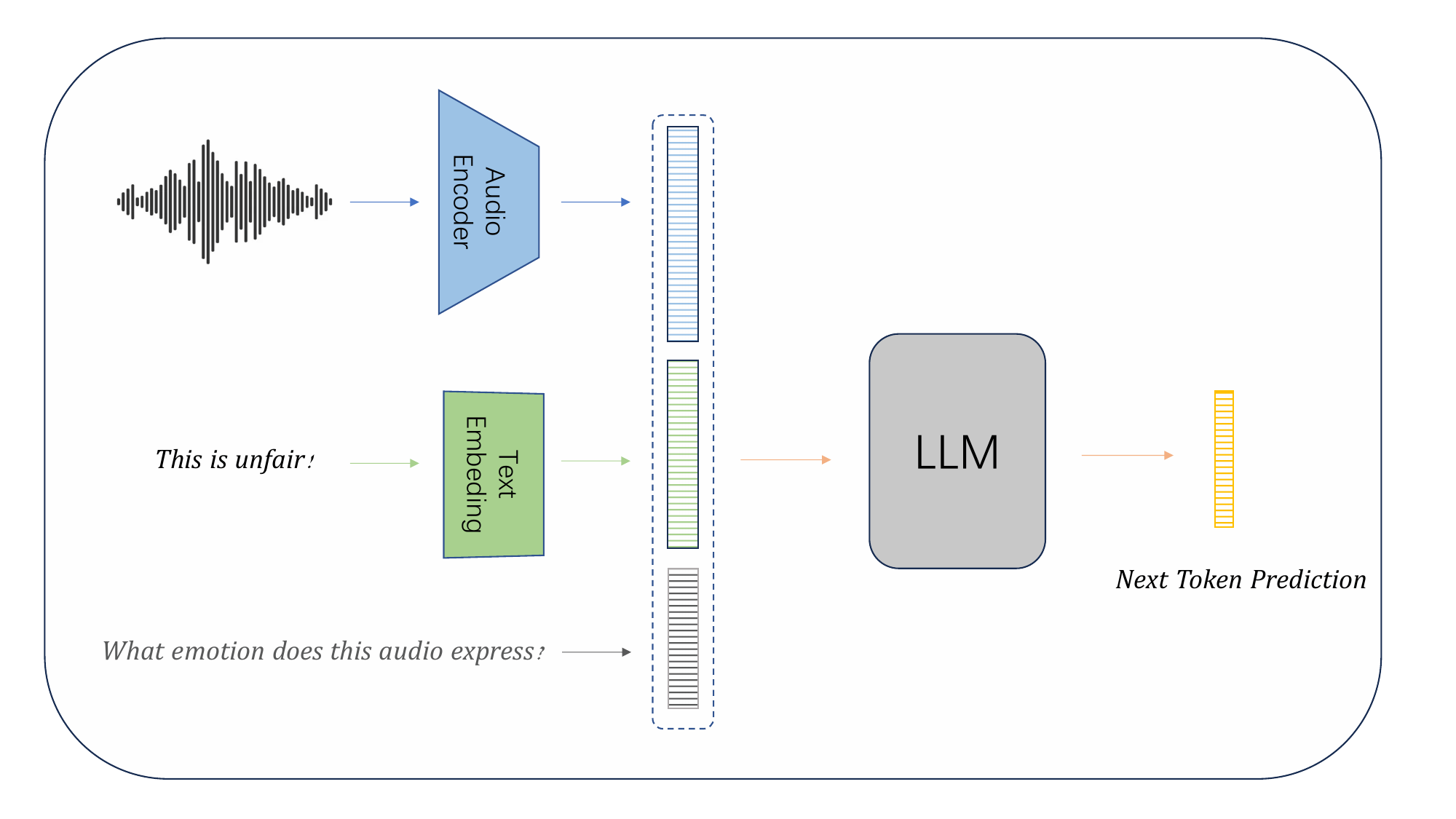}
  \caption{Illustration of Joint Audio-Text Module}
\end{figure}
In most cases, the emotion expressed by the individual in videos can be discerned through their auditory signals, encompassing not only the words but also the tone. Emotion is a product of both textual content and acoustic characteristics. For multimodal models, the crux lies in effectively integrating these modalities—textual and vocal features. Our experimental findings revealed that while features extracted using audio encoder in Section 2.2.2 and text encoder in Section 2.2.3 performed exceptionally well in their respective unimodal settings, the fusion of these two modalities based on attention mechanisms resulted in slight improvement.

We posit that sometimes there are discrepancies between audio and text representations, bringing additional competition between modalities, which potentially leads to the model under-fitting. This suggests that relying solely on feature-level fusion of cross-modal vocal-textual data might be insufficient to fully harness their complementary strengths.

To this end, we have augmented the QwenAudio\cite{chu2023qwen} framework with the inclusion of a text modality, implementing an early fusion strategy tailored for emotion-laden data. Upon training this joint audio-text framework, it evolves into a novel, unified feature extractor for audio-text modalities based on a Large Language Model, aimed at more accurately capturing and representing the intricate interplay of emotions conveyed through both audio and text. This approach endeavors to transcend the limitations of conventional multimodal fusion by fostering a deeper integration at the very onset of the processing pipeline, thereby enhancing the model's capacity to interpret emotional expressions in a multimodal context.

As shown in Figure 3, The whisper-large-v2 model is used as Audio Encoder and the decoder-only Qwen-7B model is the Large Language Model(LLM). In our methodology, audio inputs are processed through an audio feature extractor to yield audio embeddings. Concurrently, the transcribed text from the audio is transformed into text embeddings. These embeddings, along with the prompt's embedding, are then concatenated and fed into the LLM for joint training across the audio and text modalities. The high-quality dataset comprising 5,000 samples from MER2023 serves as our training data, with a notable preprocessing step that filters out instances lacking sound in an audio. Full parameter fine-tuning is adopted as the training strategy, and our experiments affirm its superiority over the LoRA method.

Despite the multitask-adapted QwenAudio model's inherent capability for speech recognition, it encounters inaccuracies when confronted with background noise or ambient sounds. Furthermore, the model's transcription output lacks punctuation, which is vital for effective emotion recognition in the textual modality as MERBench\cite{lian2024merbench} paper asserts. Consequently, furnishing the model with the correct, punctuated text aligned with the audio not only compensates for this limitation but also preserves and does not interfere with the model's innate speech recognition functionality, thereby enhancing its overall performance in multimodal emotion recognition tasks. We use our own ASR model in Section 2.4.1 to extract text from audios and punctuate the text by CT-Transformer\cite{gao2023funasr} model.

As a joint audio-text modality feature extractor, the last four layers of the large language model's decoder are employed to derive feature representations for each audio-text pair  denoted as \( \mathbf{F}_{at} \in \mathbb{R}^{N_{at} \times 4096} \).
\subsection{Noise Robustness Enhance}

\subsubsection{Noise-Robust ASR} 
\ 
\newline
\noindent Our noise-robust ASR model is improved based on Paraformer\cite{gao2022paraformer}. We optimized the ASR model from following three aspects. 

Firstly, we added noise to the training set of the ASR model for data augmentation. We employ the MUSAN\cite{musan2015} noise database as the environmental noise to expand the training set, with signal-to-noise ratios ranging from 5db to 12db. The MUSAN noise database contains various types of environmental noise, including laughter, music sounds, etc., which are very suitable for simulating noise situations in daily life. In addition, we also used the training set combined with room impulse response(RIR)\cite{szoke2019building} to produce background human voice noise to further augment our training set, with signal-to-noise ratios ranging from 10db to 30db. We train our ASR model in a 2:1:1 ratio of clean speech, ambient noise speech, and background human voice  noise speech. Secondly, we use semi-supervised method to further improve the robustness of the ASR model in noisy environments. Specifically, we use the ASR model to infer large-scale unlabeled noisy speech data, obtaining pseudo labels and beam search scores for each segment of speech. By calculating the average beam search score of each speech segment, we selected pseudo labeled data with high confidence for model self training. We iteratively trained the model until its performance did not improve. Finally, we further improved the performance of the ASR model by using a language model to avoid recognition errors. We trained an n-gram language model using a large amount of text data. Combining with Weighted Finite-State Transducers (WFST) decoding, we rescore the beams earch path of the ASR acoustic model to obtain text that is more in line with people's daily expressions. 

In this way, the performance of our ASR model can be further improved and it also has strong robustness to noisy environments. 

\subsubsection{Speech Source Separation Denoise}
\ 
\newline
\noindent For multimodal emotion recognition, audio denoising can elevate model performance significantly. Audio noise is broadly categorized into background noise and vocal noise, with separate handling of each potentially complicating the process and risking compounded audio degradation through successive denoising steps. To this end, we propose a denoising method based on MossFormer2\cite{zhao2024mossformer2}, an attention-based speech separation model capable of decomposing a noisy audio into two distinct monaural audio. 

The workflow commences by subjecting the noisy audio to noise-robust ASR model(2.4) to obtain a template text, designated as \({Text}_{temp}\). Subsequently, the noisy audio is processed through MossFormer2, yielding two separated audios. Each of the audios is then passed through the ASR model above to get transcripts \({Text}_{id0}\) and \({Text}_{id1}\).

A comparative analysis ensues, calculating the similarity between Text\_id0 and Text\_temp against Text\_temp, quantified as \({Sim}_{id0}\) and \({Sim}_{id1}\), respectively. If the absolute difference \(\lvert{Sim}_{id0}\) – \({Sim}_{id1}\rvert\) exceeds 0.1, the audio corresponding to the higher similarity score is selected; otherwise, the original audio is the chosen one. This strategy ensures a reduction in audio degradation after denoised.

The adoption of a speech source separation denoising scheme effectively mitigates the detrimental impact of noise on audio features, thereby enhancing multimodal emotion recognition model's performance.

\subsection{Data Mining and Model Ensemble}
This study proposes an innovative semi-supervised learning framework based on difference, aimed at effectively mining high confidence pseudo labeled samples from a large unlabeled dataset $D_u$, thereby promoting iterative improvement of model performance. Specifically, we first use a labeled dataset $D_l$ to train an initial weak learner through supervised learning. In order to enhance the diversity and complementarity among learners, we carefully designed four multimodal fusion models $M_1$, $M_2$, $M_3$, and $M_4$ with different feature combinations, and trained four independent weak learners $M_{l1}$, $M_{l2}$, $M_{l3}$, and $M_{l4}$ based on these models.

Subsequently, these four weak learners were synergistically applied to the unlabeled dataset $D_u$ to predict each unlabeled sample. In order to generate reliable pseudo labels, we adopted a majority voting strategy: only when the prediction results of at least three learners are completely consistent, the predicted label is adopted as the pseudo label of the sample, thus constructing the pseudo label dataset $D_w$.

Next, in order to fully utilize these newly generated pseudo labeled data, we evenly divide $D_w$ into four subsets and merge them with the original labeled dataset $D_l$ to form four extended and differentiated labeled datasets $D_{lw1}$, $D_{lw2}$, $D_{lw3}$, $D_{lw4}$.

Subsequently, the four extended datasets were used as training sets for supervised and refined training of models M1 to M4, resulting in four more powerful learners $M_{lw1}$, $M_{lw2}$, $M_{lw3}$, $M_{lw4}$.

We used the updated learner to vote on the remaining unlabeled data to generate new pseudo labeled samples and repeat the training and enhancement steps above for N iterations. As the number of iterations increases, the model's performance improve gradually driven by the continuous enrichment of data.

After N iterations, we obtained 4 strong learners. These learners adopt different strategies in feature extraction and different combinations of datasets during training, resulting in significant differences in model characteristics and knowledge representation.

Based on four strong learners, we designed a multimodel ensemble strategy, as shown in Figure 4. We rank four models from high to low according to the indicators on the test set, and then count the mode(most common categories) of the output results of multiple models. If the number of modes is 1, we choose the mode as the voting result, otherwise we ignore the last model's result, and repeat the above steps until the mode is unique.
\begin{figure}[htbp]
\centering
\subfigure[]{
\includegraphics[height=3cm]{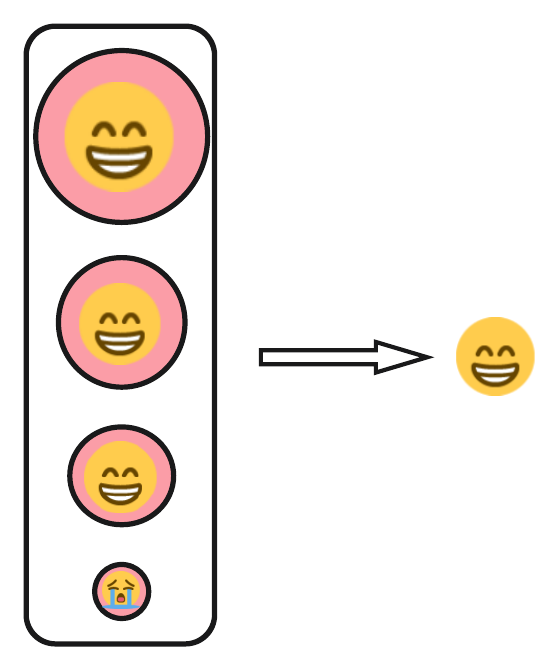}
\label{(a)}}
\quad
\subfigure[]{
\includegraphics[height=3cm]{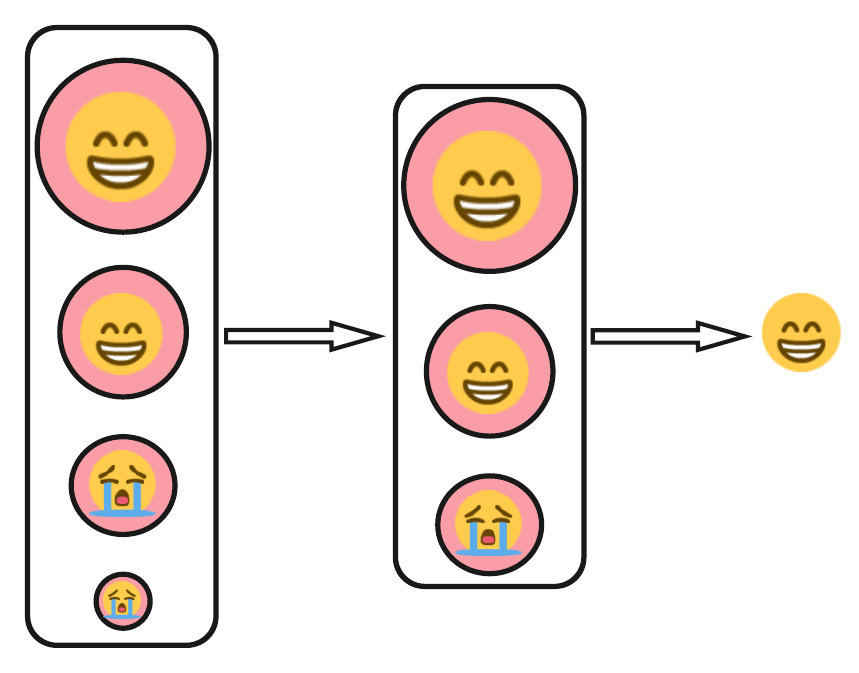}
\label{(b)}}
\caption{MuiltiModel Ensemble Strategy}
\end{figure}
\section{Experiments and Analysis}
\subsection{Dataset}
The data details of MER-SEMI and MER-NOISE tracks are shown in Table 1. 
\begin{table}[H]
\caption{\textbf{MER2024 Dataset Distribution}}
\centering
\resizebox{0.5\linewidth}{!}{
\begin{tabular}{cc}%
\toprule
MER2024 &samples(labeled/whole) \\
\midrule
Train\&Val & 5030 \\
MER-SEMI & 1169/115595 \\
MER-NOISE & 1170/115595 \\
\bottomrule
\end{tabular}}
\end{table}
This dataset includes 5030 carefully labeled data samples and 115595 unlabeled data samples, which together form the foundation of the training and validation stages. Participants can fully utilize this data to optimize and improve the performance of their models. In the testing phase, in order to encourage participants to pay more attention to the generalization ability of the model, it is required to make predictions on all submitted models on a sample set of 20000. However, for the MER-SEMI track, we only evaluated 1169 samples; For the MER-NOISE track, 1170 noisy samples were evaluated to comprehensively test the performance of the model in complex noise environments.

We have trained two vision feature extractors, based on VIT-Base and InternVIT-6B respectively. Training VIT-Base encoder includes pre-train and fine-tune stages. In pre-train stage, we collected CelebA\cite{liu2015deep}, VoxCeleb2\cite{chung2018voxceleb2}, LFW\cite{huang2008labeled}, wideface\cite{yang2016wider}, FFHQ\cite{karras2019style}, IMDB-WIKI\cite{rothe2015imdb}, CH-SIMS\cite{yu2020ch}, CMU-MOSI\cite{zadeh2018multi}, and RAF-DB \cite{li2017reliable}datasets, with a total number of 9M. After filtering out blurry faces, incomplete faces, and large profile faces, the remaining 8M is used as our pre-train dataset. In the fine-tune stage, we use the labeled dataset ExpW\cite{zhang2015learning} and RAF-DB dataset related to emotion and labeled data of MER2024. InternVIT-6B shares the same fine-tune data.

\subsection{Ablation Study}
For ViT-Base, we choose mmpre-train\cite{2023mmpre-train} framework for large-scale pre-training and fine-tune it by adding a classification header. In the pre-train stage, the parameters are set as follows: input resolution 224, batch size 1024, initial learning rate 2e-4, maximum training rounds 300, mixed precision training enabled, gradient accumulation 4, and training on 8*A800. In the finuetune stage, the final learning is 2e-3, the maximum number of training rounds is 100, and the other settings are the same as pre-train stage. 

For InternViT-6B, the experimental settings are as follows, batch size 512, input resolution 224, learning rate 2e-5, optimizer SGD, attenuation parameter 0.9, maximum training rounds 50. 
\begin{table}[H]
    \caption{Emotion ViT fine-tune}
    \label{table}
    \setlength{\tabcolsep}{4pt}
    \label{ViT-Base}
    \resizebox{1.0\linewidth}{!}{
    \begin{tabular}{cccccc} 
        \hline
        &\multicolumn{3}{c}{ViT-Base}& \\
        Adamw\cite{loshchilov2017decoupled} & Single & Multi & 8classes & 3classes & $WAF_{MER2023-multi}(\uparrow)$ \\
        \hline
        \Checkmark & \Checkmark& & &\Checkmark &61.85 \\ 
        \Checkmark & \Checkmark& & \Checkmark &  &62.51 \\
        \Checkmark & &\Checkmark & & \Checkmark &63.45 \\
        \Checkmark & & \Checkmark & \Checkmark & & \textbf{65.07} \\
        \hline
    \end{tabular}
    \label{InternViT-6B}
    \begin{tabular}{cccc}
        \hline 
        &\multicolumn{2}{c}{InternViT-6B}& \\
        25-47 Blocks&0-25 Blocks& $CLIP_{proj}$ & $WAF_{MER2023-multi}(\uparrow)$ \\
        \hline
        \Checkmark& \Checkmark &\Checkmark &63.66 \\ 
         & \Checkmark & &64.68 \\
        \Checkmark & & &66.63 \\
          & &\Checkmark &\textbf{67.30} \\
        \hline
    \end{tabular}}
\end{table}
Table 2 shows the details of ViT-Base and InternViT-6B in fine-tuning. The model on the left is ViT Base, which uses the Adamw optimizer. "single" represents a single image as the input to the training set, as shown in Figure 2 (a). "Multi" represents the use of multi image stitching as the input to the video data in the training set, as shown in Figure 2 (b). Non video data still uses a single input. "8classes" represents 8 categories when fine-tuning, which are "angry", "disgust", "fear", "happy", "sad", "surprise", "neutral", and "worried". "3classes" represents 3 categories, which are "positive", "negative" and "neutral". The model on the right is the experimental result of fine-tuning InternViT-6B. 25-47 Blocks indicate that fine-tuning is released for training on layers 25-47, while the rest are frozen. 0-25 Blocks indicate that layers 0-25 are released for training, while the rest are frozen. $CLIP_{proj}$ indicates that only the projector layer is released. Comparative experiments on the results of different classification categories of data training are shown on Table 2's left, which shows that the more classification categories, the more the model improves. In addition, it is also verified that multi-frames input can improve the model's video recognition capabilities.

The results show that full-parameter training has the worst effect, and only a few parameters of $CLIP_{proj}$ are released, which has the best effect on downstream tasks.

In addition, we conducted experiments to verify the performance of the joint Audio-Text feature extractor. The training dataset is labeled data over 3k in MER2023. The test dataset includes test dataset in MER2023's multi-label track and the test dataset in both tracks of MER2024. As shown in Table 3, Audio is the audio feature extracted from fine-tuned $Chinese\_Hubert\_large$, Text refers to the text feature extracted from fine-tuned $Baichua\_13B$, Joint Audio-Text is the joint Audio-Text feature based on enhanced QwenAudio ($Qwen\_JAT$), and 'Denoise' represents denoising preprocessing. 

From the table, we can find that Joint Audio-Text feature fused with $Baichuan\_{13B}_{ft}$ and $Chinese\_Hubert\_large_{ft}$ reached 0.8606 score in MER2024-semi track. Tri-feature fusion can maximize the performance of audio and text modality. In addition, using Joint Audio-Text features after audio denoising can simultaneously improve our model's performance in both tracks of MER2024.However, due to the reduction of the speech sampling rate after denoising (16k to 8k), the feature quality of $Chinese\_Hubert\_large_{ft}$ (pre-trained on audios sampling rate 16k) decreased, which cause the score in noise-track dropped after the tri-feature fusion.
\begin{table}[H]
\caption{\textbf{Comparison results around Joint Audio-Text modality feature}}
\centering
\resizebox{0.9\linewidth}{!}{
\begin{tabular}{ccccccc}
\toprule
Audio &Text &Joint Audio-Text& Denoise &\(WAF_{MER2023-multi}(\uparrow)\)&\(WAF_{semi}(\uparrow)\)&\(WAF_{noise}(\uparrow)\) \\
\midrule
\Checkmark & & & &73.34 & — & — \\
 &\Checkmark & & &62.20 & — & — \\
& &\Checkmark & &75.36 &85.43 & 74.23 \\
& &\Checkmark &\Checkmark &75.11 &85.82 & \textbf{77.10} \\
\Checkmark &\Checkmark & & &75.31 &82.30 & —\\
\Checkmark & &\Checkmark & &75.68 &85.48 & —\\
 &\Checkmark &\Checkmark & &75.54 &85.51 & —\\
\Checkmark &\Checkmark &\Checkmark & &\textbf{76.22}&\textbf{86.06}& 74.78\\
\Checkmark &\Checkmark &\Checkmark &\Checkmark &75.83 &85.98 & 76.61\\
\bottomrule
\end{tabular}}
\end{table}

Next, we conduct features fusion experiments extracted from all our modal feature encoder based on MER2024 labeled dataset sized 5030. The Table 4 shows the comparison of different modality feature groups with the baseline in the multi-modal fusion model. It's clear that our fine-tuned audio encoder (${Chinese\_Hubert\_larg}e^{\ast}$), fine-tuned text encoder (${Baichuan\_13B}^{\ast}$), $Emotion ViT$(collective name for $vit\_base\_emo$ and $internvit\_6B\_emo$) and $Qwen\_JAT$ all have significantly improved the overall performance of multi-modal models respectively. 
\begin{table}[H]
\caption{\textbf{Comparison results in MER2023 and MER2024 test
dataset of different multi-modal feature groups.}}
\centering
\renewcommand\arraystretch{2.0}
\resizebox{0.9\linewidth}{!}{
\begin{tabular}{ccccccc}
\toprule
A &T & V & Joint A-T & \(WAF_{MER2023-multi}(\uparrow)\)&\(WAF_{semi}(\uparrow)\)& \(WAF_{noise}(\uparrow)\) \\
\midrule
$Chinese\_Hubert\_large$ & $Baichuan\_13B$ & $CLIP\_large$ & — & 82.17 & 84.45 & — \\
$Chinese\_Hubert\_large$ & $Baichuan\_13B$ & $internvit\_6B\_emo$ & — & 82.76 & — & — \\
$Chinese\_Hubert\_large$ & $Baichuan\_13B$ & $vit\_base\_emo$ & — & 83.29 & — & — \\
$Chinese\_Hubert\_large$ & ${Baichuan\_13B}^{\ast}$ & $vit\_base\_emo$ & — & 84.43 & — & — \\
${Chinese\_Hubert\_larg}e^{\ast}$ & ${Baichuan\_13B}^{\ast}$ & $vit\_base\_emo$ & — & 85.31 & 86.94 & — \\
— &${Baichuan\_13B}^{\ast}$ & $vit\_base\_emo$ & $Qwen\_JAT$ & — & 87.86 & 79.54 \\
— & ${Baichuan\_13B}^{\ast}$ & $vit\_base\_emo$ & ${Qwen\_JAT}_{denoised}$ & — & \textbf{88.25} & \textbf{82.44} \\
\bottomrule
\end{tabular}}
\end{table}
Specifically, the encoder groups including ${Baichuan\_13B}^{\ast}$, $vit\_base\_emo$ and $Qwen\_JAT$ increase the $WAF$ by up to $3.41\%$ compared to the baseline, which is a very considerable improvement.

Moreover, our denoise preprocess method for audios before $Qwen\_JAT$ showed outstanding results on the MER2024-NOISE track, with a WAF increase of $2.9\%$, proving the effectiveness and robustness of our approach. What’s even more gratifying is that in the MER2024-SEMI track, our denoise method also brings a $0.39\%$ increasement.

In order to amplify the diversity among models, we designed four distinct feature combination schemes. Just as shown in Figure 5, through multi-turns data mining , we elevated the Weighted Average F1-score (WAF) of the four individual models above 0.88, with the optimal combination approaching 0.89. Finally, by ensemble voting on these four models with remarkable differences, we effectively raised the WAF to 0.9001.
\begin{figure}[h]
  \centering
  \includegraphics[width=0.4\textwidth]{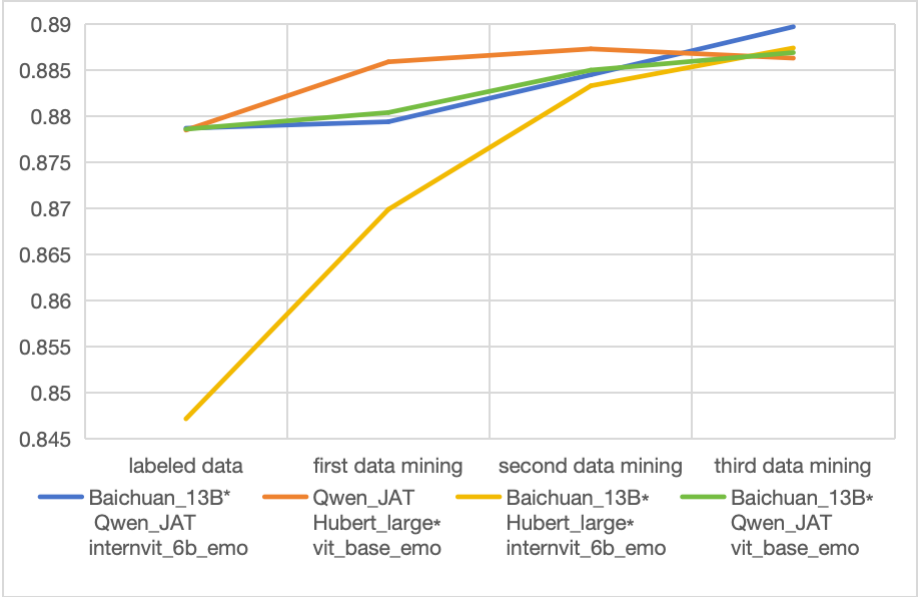}
  \caption{The trend of model indicators with data mining iteration.}
\end{figure}
\section{Conclusion and Limitations}
In this paper, we present our solutions for emotion recognition in the sub-challenges of MER2024. Firstly, We launched our Emotion ViT based on large-scale data pre-training and fine-tune, a vision feature extractor adapted to emotion recognition tasks. To mitigate the modal competition issue between audio and text, we adopt an early fusion strategy based on a large language model, where joint training of audio and text is conducted initially. And the joint Audio-Text modal feature will be late-fused with other unimodal features. In order to solve the problems of data insufficiency and class imbalance, We use multiple turns of multi-model voting for data mining. Moreover, to enhance the quality of audio features, we employ speech source separation to preprocess audios. The efficacy of our methods has been demonstrated through outstanding performances in both MER2024-SEMI and MER2024-NOISE, achieving scores of 0.9001 and 0.8383 respectively.

Our research into the competition among visual modality and other modalities is insufficient, and furthermore, the ambiguity inherent in Chinese text poses a particular challenge for emotion recognition. These issues above may need the utilization of large language models for resolution. In the future, we will investigate emotion recognition based on Multimodal Large Language Models.




\end{sloppypar}
\bibliography{references}
\end{document}